\documentclass[12pt,linenumbers]{article}
\usepackage[utf8]{inputenc}
\usepackage{amsmath}
\usepackage{sectsty}
\usepackage{graphicx}
\usepackage{amssymb}
\usepackage{caption}
\usepackage{xcolor} 
\usepackage[normalem]{ulem} 
\usepackage[sorting=none,style=nature,citestyle=nature,doi=false ,isbn=false,url=false,eprint=false, autocite = superscript,backend=biber]{biblatex}
\usepackage{times}
\usepackage{float}
\usepackage{afterpage}

\title{Dominant end-tunneling effect in two distinct Luttinger liquids coexisting in one quantum wire}

\author{Henok Weldeyesus,$^{1,\dagger}$ Pedro M.T. Vianez,$^{2,3,\dagger}$ Omid Sharifi Sedeh,$^{1,\dagger}$ \\ Wooi Kiat Tan,$^{2}$ Yiqing Jin,$^{2}$ Mar\'ia Moreno,$^{2,4}$ Christian P. Scheller,$^{1}$ \\ Jonathan P. Griffiths,$^{2}$ Ian Farrer,$^{5}$ David A. Ritchie,$^{2}$ \\ Dominik M. Zumb\"uhl,$^{1,\ast}$ Christopher J.B. Ford,$^{2,\ast}$ Oleksandr Tsyplyatyev$^{6,\ast}$\\
\\
\normalsize{$^{1}$Department of Physics, University of Basel, Klingelbergstrasse 82,} \\ 
\normalsize{4056 Basel, Switzerland}\\
\normalsize{$^{2}$Department of Physics, Cavendish Laboratory, University of Cambridge,}\\ \normalsize{Cambridge, CB3 0HE, UK}\\
\normalsize{$^{3}$ Los Alamos National Laboratory, Los Alamos, New Mexico 87545, USA}\\
\normalsize{$^{4}$Departamento de F\'isica Aplicada, Universidad de Salamanca,}\\
\normalsize{Plaza de la Merced s/n, 37008 Salamanca, Spain}\\
\normalsize{$^{5}$Department of Electronic and Electrical Engineering, University of Sheffield,}\\
\normalsize{Sheffield, S1 3JD, UK}\\
\normalsize{$^{6}$Institut f\"ur Theoretische Physik, Universit\"at Frankfurt, Max-von-Laue Stra{\ss}e 1,}\\
\normalsize{60438 Frankfurt, Germany}\\
\\
\normalsize{$^\dagger$These authors contributed equally to this work}\\
\normalsize{$^\ast$To whom correspondence should be addressed; E-mail:}\\ \normalsize{tsyplyatyev@itp.uni-frankfurt.de, cjbf@cam.ac.uk, dominik.zumbuhl@unibas.ch}\\
}

\date{}

\begin{document}
\maketitle

\newpage
\begin{abstract}
Luttinger liquids occupy a notable place in physics as one of the most understood classes of quantum many-body systems. The experimental mission of measuring its main prediction, power laws in observable quantities, has already produced a body of exponents in different semiconductor and metallic structures. Here, we combine tunneling spectroscopy with density-dependent transport measurements in the same quantum wires over more than two orders of magnitude in temperature to very low temperatures down to $\sim$40\,mK. This reveals that, when the second 1D subband becomes populated, the temperature dependence splits into two ranges with different exponents in the power-law dependence of the conductance, both dominated by the finite-size effect of the end-tunneling process. This result demonstrates the importance of measuring the Luttinger parameters as well as the number of modes independently through spectroscopy in addition to the transport exponent in the characterization of Luttinger liquids. This opens a pathway to unambiguous interpretation of the exponents observed in quantum wires.
\end{abstract}

\section*{Introduction}
Out of all many-body phenomena in quantum physics, Luttinger liquids occupy a paradigmatic place as one of the most established cases of interactions changing entirely the basic properties of the underlying particles. Such a strongly correlated state is realized in one-dimensional (1D) systems and is theoretically described by the hydrodynamic Tomonaga-Luttinger theory \cite{Tomonaga50,Luttinger63,Haldane81b}. On the microscopic level, the many interacting particles form density waves already at low energy, producing interaction-dependent power laws in the correlation functions \cite{Schoenhammer92,Voit93} and, therefore, in various observables, which is one of the hallmark predictions of Luttinger-liquid physics. It was more recently generalized to the whole, usually nonlinear, energy band \cite{Imambekov09,Imambekov09p,Jin19}. The other signature prediction of Luttinger liquids is separation of the spin and charge degrees of freedom for particles with spin, i.e., the velocities of spin and charge-density waves are different. This was recently generalized to the whole nonlinear band in \cite{Tsyplyatyev22,Vianez21}.

\begin{figure}[h!]
	\centering
	\includegraphics[width=0.75\linewidth]{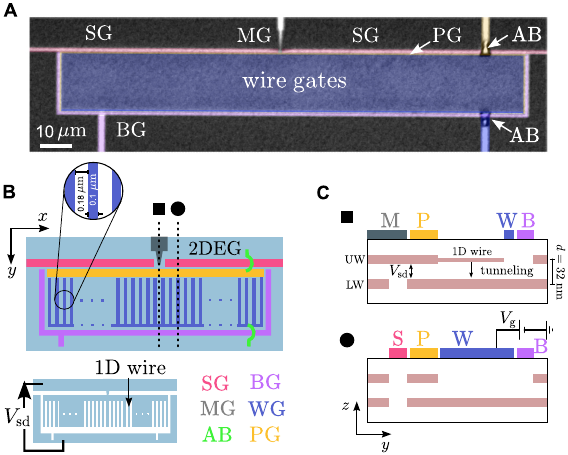}
	\caption{Schematics of the device. {\bf A} Optical micrograph of the device, showing the very regular array of wire gates as a uniform blur in the center. The air bridges provide electrical connections to the p and wire gates.
		{\bf B} Top view with the upper well (UW) and the electrostatic gates (color-coded). A narrow region (p-region) in the upper well remains 2D and is covered by a gate `p' (labeled PG) to allow tuning of its density.
		Lower panel shows depleted (white) and non-depleted (light blue) regions of the upper 2DEG after all voltages are set.
		{\bf C} Side views of the double-well structure, showing  where tunneling from a wire occurs to the lower 2DEG ($\blacksquare$), and a region between wires ({\Large$\bullet$}), corresponding to the dashed lines in {\bf B}.  The centers of the upper well (UW) and lower well (LW) are separated by $d=32$\,nm. The UW 2DEG beneath the wire gate is formed into an array of 1D quantum wires by the negative voltage on the wire gate $V_{\rm g}$, and $V_{\rm sd}$ is the source-drain voltage between two wells. Other gates: AB is an air bridge, BG is the barrier gate allowing current to flow only by tunneling; SG is the split gate depleting both wells and MG is the mid-gate, injecting current only into UW.} 
	\label{fig:device}
\end{figure}

The experimental challenge of observing the Luttinger-liquid behavior was first approached by measuring the power law in transport experiments, where the tunneling conductance vanishes at small voltages (called the zero-bias anomaly or ZBA) due to the vanishing of the density of states for still gapless density-wave excitations at the Fermi energy \cite{Tomonaga50,Luttinger63}. This was observed in carbon nanotubes \cite{Bockrath99,Yao99,Bachtold01}, in NbSe$_3$ \cite{Slot04} and MoSe \cite{Venkataraman06} nanowires, in GaAs 2D electron gases (2DEG) with electrons localized at the edge by means of the quantum-Hall effect \cite{Grayson98}, and later in quantum wires formed electrostatically \cite{Auslaender02,Jompol09}. However, interpretation of the observed exponents in terms of the Luttinger-liquid theory was always based on less reliable theoretical assumptions about the interaction strength that is open to different interpretations since different tunneling mechanisms such as bulk \cite{Kane97,Altland99}, end \cite{Eggert97,Kane97}, and through-a-barrier \cite{Kane92} tunneling processes predict different exponents, and are nearly impossible to discriminate between without independent knowledge of the Luttinger-liquid parameters. Separately, the spin-charge separation was observed as two (rather than one) linear modes with different velocities around the Fermi energy using angle-resolved photoemission spectroscopy in a strongly anisotropic organic conductor TTF-TCNQ \cite{Zwick98}, in a high-$T_{\rm c}$ superconductor ${\rm SrCuO_2}$ \cite{Kim96} and also by using magnetotunneling spectroscopy in GaAs heterostructures \cite{Auslaender02,Jompol09}. It was also measured in time-of-flight experiments as two wavefronts propagating with different velocities in cold ${\rm ^6Li}$ atoms on an optical lattice \cite{Hilker17,Salomon19,Vijayan20} and in chiral quantum-Hall states in GaAs \cite{Hashisaka17}.  Such spectroscopy, in contrast to the power-law measurements, gives independent experimental access to the interaction parameters directly.

Here, we choose a semiconductor wire to 2DEG tunneling setup \cite{Jompol09} to measure transport and spectroscopy in the same quantum wire simultaneously using the magnetotunneling technique. A highly optimized and well-filtered dilution refrigerator gives us access to a wide temperature range from about 5\,K down to 8\,mK.
By varying the electronic density systematically, we find one or sometimes two Luttinger-liquid exponents in over two decades of temperature.
Then, we measure spectroscopy for each electronic density at low temperature to extract the microscopic parameters of the Luttinger liquid in our wires. By comparing our directly obtained exponents with the predictions of the Luttinger-liquid theory, we find that the experimental values are an order of magnitude larger than the theoretical ones for the bulk-tunneling transport channel but are close to the predicted values for the end-tunneling regime. Therefore, we associate the appearance of the second exponent at higher densities with the occupation of the second 1D subband, which is accessible in semiconductor wires and is indicated by the appearance of the second Fermi point in the spectroscopic data. This measurement demonstrates the coexistence of two fairly independent Luttinger liquids with two different sets of Luttinger parameters in the same wire, which could offer a new setup for Coulomb-drag experiments in 1D \cite{Nazarov98,Pustilnik03,Yamamoto06,Laroche14}. This result shows that the challenge of measuring one of the main fundamental predictions of Luttinger liquids (bulk power laws) in semiconductor wires still remains open, and 
raises the question of whether the `bulk' exponents observed in some carbon nanotube experiments \cite{Bockrath99,Yao99,Bachtold01} are also due to a similar finite-size effect, since they are so large that it requires the assumption of very strong interaction strength to interpret them as the bulk effect.

\section*{Results}
\subsection*{Transport exponent}
In our experiment, the differential conductance $G$ is measured in an out-of-wire tunneling setup in a GaAs/Al$_{0.33}$Ga$_{0.67}$As double-well heterostructure in Fig.~\ref{fig:device}, with a finite, in-plane magnetic field applied perpendicular to the wires. 
\begin{figure}
	\centering
	\includegraphics[width=0.95\linewidth]{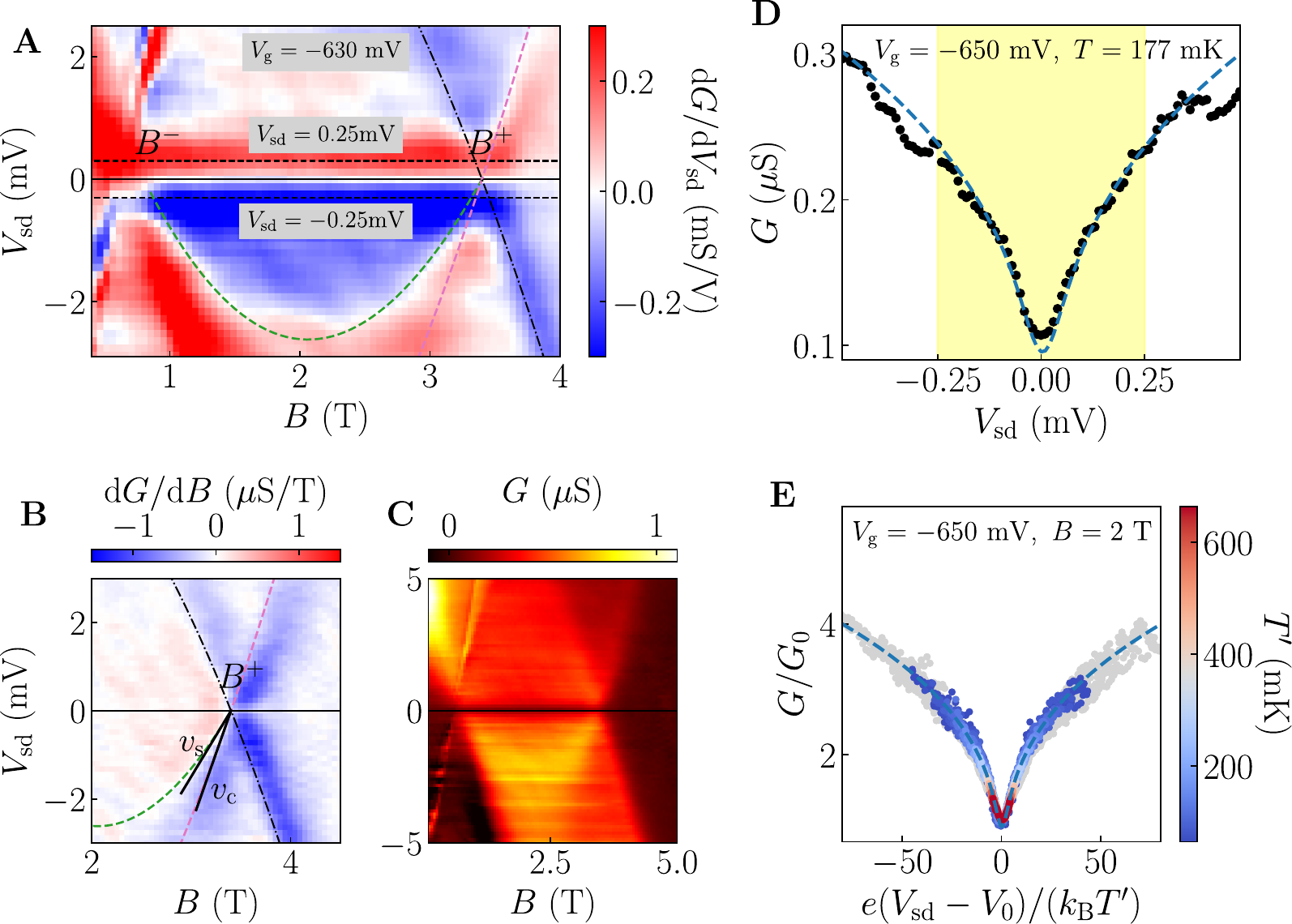}
	\caption{Spectroscopic maps and universality of conductance at low energy. {\bf A} Tunneling conductance $G(B, V_{\rm sd})$ in the single-subband regime for $V_{\rm g}=-630$\ mV at a lattice temperature of $8$\,mK presented as derivative of $G$ with respect to the voltage $V_{\rm sd}$, ${\rm d}G/{\rm d}V_{\rm sd}$. The black dashed lines around the $V_{\rm sd}=0$ line mark the extent of the linear region around the Fermi energy, $V_{\rm sd}=\pm 0.25$~mV, in which the conductance is mostly independent of magnetic field (and momentum). The green and pink dashed lines on all panels mark the dispersions of the spin and charge Fermi seas, respectively. The black dash-dotted line marks the dispersion of the 2DEG in the bottom well measured by the Fermi edge of the quantum wire. The $B^\pm$ points correspond to the $\pm k_{\rm F}^{\rm 1D}$ points of the 1D electrons. The details of fitting the features are given in the text. {\bf B} Derivative of $G$ with respect to the magnetic field $B$, ${\rm d}G/{\rm d}B$ around the point labeled $B^+$. The two solid lines mark the spin ($v_{\rm s}$) and charge ($v_{\rm c}$) velocities around this point. {\bf C} Map of the measured tunneling conductance $G(B, V_{\rm sd})$ showing how $G$ vanishes at $V_{\rm sd}=0$.  {\bf D} Voltage cut at $B=2$\,T and $T=177$\,mK for $V_{\rm g}=-650$\,mV. The yellow rectangle marks the linear regime $|V_{\rm sd}|<0.25$\,mV. {\bf E} Rescaled conductance, $G(eV_{\rm sd}/k_{\rm B}T^\prime)/G_0$, in the linear regime in the 8 to 670\,mK range, in which the electronic temperature $T^\prime$ is used to take into account the electron-phonon decoupling at $T<65$\,mK. The colors of the points correspond to the temperatures shown in the bar on the right, except that gray is used for points outside the linear regime $|V_{\rm sd}|<0.25$\,mV. The data are measured in the single-subband regime at $V_{\rm g}=-650$\,mV and $B=2$\,T and the dashed-blue line is Eq.\ (\ref{eq:G_finite_T}) with $\alpha=0.36$ in {\bf D} and {\bf E}.
	} \label{fig:conductance_1subband}
\end{figure}

We start by setting the wire-gate voltage to $V_{\rm g}=-630$\,mV, close to pinch-off, so that only a single 1D subband in the wires in the upper well is expected to be populated. The conductance map for a wide range of interlayer voltages $V_{\rm sd}$ and magnetic fields $B$ is presented in Fig.\ \ref{fig:conductance_1subband}C; in Figs.\ \ref{fig:conductance_1subband}A and B derivatives of the same data with respect to $V_{\rm sd}$ and $B$ are shown to help visualize different features. The contribution to the signal from the wires shows two separate features, both with parabolic dispersions away from $V_{\rm sd}=0$, and a zero-bias anomaly (ZBA) around the $V_{\rm sd}=0$ line, which is almost independent of $B$ over a wide range. The former is the nonlinear effect of the spin-charge separation of the Fermi sea due to Coulomb interactions \cite{Vianez21}, which we have shown can be described by two parabolae using the Fermi-Hubbard model \cite{Tsyplyatyev22}, and the latter is the linear effect of the vanishing density of states at the Fermi level, which can be described by the Tomonaga-Luttinger model \cite{Tomonaga50,Luttinger63}. The boundary between these two regimes can be found by inspecting the conductance maps, e.g., $\left|V_{\rm sd}\right|=0.25$\,mV in Fig.~\ref{fig:conductance_1subband}A. In this work we are mostly interested in the low-energy physics, so we focus on the ZBA.
 
One of the predictions of the Tomonaga-Luttinger model is that the conductance does not depend on voltage $V_{\rm sd}$ and temperature $T$ independently but is given by a universal scaling curve of their ratio \cite{Fisher85,Grabert85},
\begin{equation}
  G(V_{\rm sd},T)=A T^\alpha \cosh\left(\frac{eV_{\rm sd}}{2k_{\rm B}T}\right) \left|\Gamma\left(\frac{1+\alpha}{2}+\frac{{\rm i} eV_{\rm sd}}{2\pi k_{\rm B}T} \right)\right|^2,\label{eq:G_finite_T}
\end{equation}
where $A$ is a temperature- and voltage-independent constant, $\alpha$ is a transport exponent predicted by the Tomonaga-Luttinger model at $T=0$ that depends on the interaction strength, $\Gamma(x)$ is the gamma function, $k_{\rm B}$ is the Boltzmann constant, and a parameter describing the voltage division between two tunnel junctions is not required since in our setup almost all the voltage drops in across the tunnel barrier between two quantum wells. To check this prediction, we measure voltage cuts in the whole map in Fig.~\ref{fig:conductance_1subband}A-C at a fixed magnetic field around the Fermi point (where the signal is strongest) slowly increasing the temperature step-wise from the base temperature of $8$\,mK to $600$\,mK to ensure sample thermalization throughout the process. The temperature is controlled with a heater on the flange of the mixing chamber and measured with a RuO$_2$ thermometer. Further details on the measurement setup are given in Methods.

The results are presented as a superposition of all the measured voltage cuts at the same magnetic field of $B=2$\,T for each temperature over a wide range as a function of $eV_{\rm sd}/k_{\rm B} T^\prime$ in Fig.~\ref{fig:conductance_1subband}E. An effective electron temperature $T'=\sqrt[3]{T_0^3+T^3}$ \cite{Jompol09,Casparis12} with an electron saturation temperature $T_0=65$\ mK was used in place of $T$ to take into account the saturation of the data at $T\lesssim T_0$, which we interpret as an effect of electron-phonon decoupling. 
For low voltages, the curves collapse on to the same universal curve as predicted by Eq.~(\ref{eq:G_finite_T}). However, they all become non-universal beyond a certain voltage that marks a crossover to the nonlinear regime. There the conductance needs rather to be described by a different, nonlinear model \cite{Imambekov09,Imambekov09p,Tsyplyatyev14,Schmidt10,Tsyplyatyev15,Tsyplyatyev16,Moreno16,Jin19,Vianez21,Tsyplyatyev22,vianez_book} dominated by the spin-charge splitting of the Fermi sea \cite{Vianez21,Tsyplyatyev22,vianez_book}, which is characterised by an essential dependence on magnetic field (i.e., on the momentum of the collective modes) and the absence of the particle-hole symmetry and of the universal conductance scaling. To assess the crossover point to the nonlinear regime in the voltage domain quantitatively, we select a single voltage cut at an intermediate temperature and fit it with Eq.~(\ref{eq:G_finite_T}) using the exponent $\alpha$ as a fitting parameter in Fig.~\ref{fig:conductance_1subband}D. In such a fit, we use the particle-hole symmetry of the linear Tomonaga-Luttinger model to restrict the fitting window at low voltages: the points where the amplitudes of the signal for positive and negative voltages $\pm V_{\rm sd}$ start to deviate from each other marks the crossover, giving us $V_{\rm sd}=0.25$\,mV as the range of validity of the low-energy regime. Note that the data in Fig.~\ref{fig:conductance_1subband}E was measured in the single-subband regime at a density of $n_{\rm 1D}=40\,\mu\mathrm{m}^{-1}$, see Supplementary Fig.\ 4, corresponding to a chemical potential $\mu=2-3$\,meV that can be seen directly in the data in Fig.~\ref{fig:conductance_1subband}A as $e$ times the negative voltage needed to reach the bottom of the green dashed parabola. For different densities in the wires, the crossover point is different and is generally expected to be smaller than $V_{\rm sd}=0.25$\,mV for lower densities.
\begin{figure}
	\centering
	\includegraphics[width=0.95\linewidth]{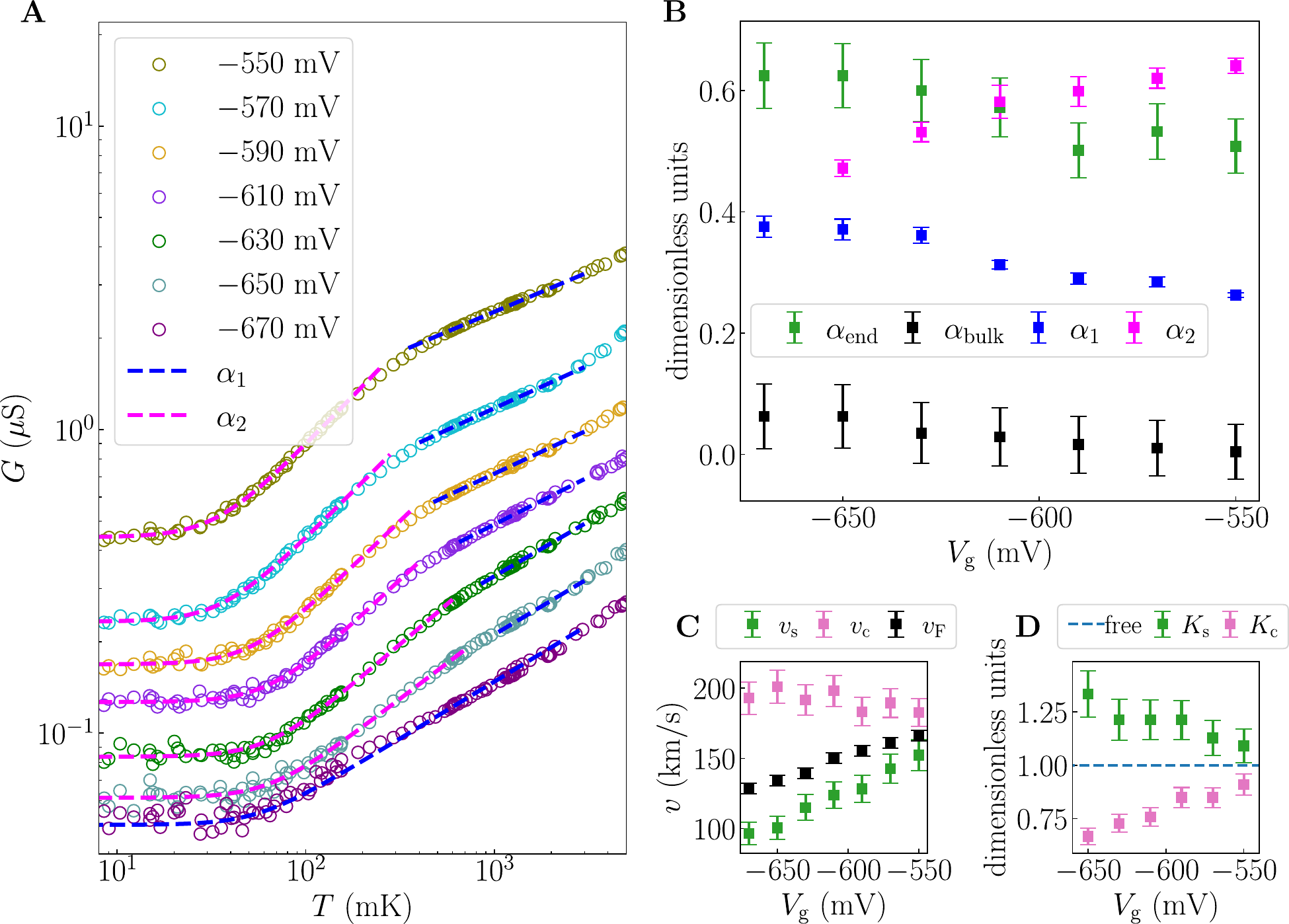}
	\caption{Temperature dependence of conductance and dependence of the Luttinger parameters on electron density. {\bf A} Conductance at $V_{\rm sd}=0$ as a function of temperature on a logarithmic plot for the gate voltages $V_{\rm g}$ given in the legend. The blue and magenta dashed lines are the power-law fits giving the values of the exponents in {\bf B}. The details of the fitting procedure are given in the text. {\bf B} The values of two exponents $\alpha_{\rm 1}$ (blue squares) and $\alpha_{\rm 2}$ (magenta squares) as a function of $V_{\rm g}$ extracted from the conductance data in {\bf A} with the error bars showing the rms error in the fit. The bulk-transport exponent $\alpha_{\rm bulk}$ (black squares), the end-transport exponent $\alpha_{\rm end}$ (green squares), and their error bars are evaluated for the Luttinger parameters in {\bf D} using Eq.~(\ref{eq:alpha_bulk}) and Eq.~(\ref{eq:alpha_end}), respectively. {\bf C} The velocities of excitations of spin ($v_{\rm s}$, green squares) and charge ($v_{\rm c}$, pink squares) extracted from the spectroscopic maps, e.g., Fig.~\ref{fig:conductance_1subband}A, as the linear slopes around the $B_+$ point, the Fermi velocity $v_{\rm F}$ extracted from the distance between the $B_\pm$ points, and the error bars indicate the range of values that give an acceptable fit. {\bf D} The Luttinger parameters for spin ($K_{\rm s}$, green squares), charge ($K_{\rm c}$, pink squares), and their error bars obtained from the data in {\bf C} using $K_\nu=v_{\rm F}/v_\nu$. The blue dashed line is the non-interacting limit of these parameters, $K_{\rm s,c}=1$.} 
	\label{fig:conductance_exponents}
\end{figure}
  
Now we measure the zero-bias conductance as a function of temperature, over a wide range of about three decades, and for a range of different $V_{\rm g}$ corresponding to different densities $n_{\rm 1D}$ (see Supplementary Fig.~4) in the middle of the linear regime. The result is presented on a log-log scale in Fig.\ \ref{fig:conductance_exponents}A. According to Eq.~(\ref{eq:G_finite_T}), the Luttinger-liquid exponent $\alpha$ should be directly visible as a straight line in this figure. What we in fact observe is two different exponents in the range $\alpha=0.3-0.6$, summarized by the blue and magenta points in Fig.~\ref{fig:conductance_exponents}B. In extracting the exponents, we exclude temperatures $T>1-3$\,K from the analysis since the thermal energy is already in the nonlinear regime corresponding to $eV_{\rm sd}\gtrsim 0.25$\,meV. For the lowest temperatures of $T<35-65$\,mK, the signal saturates within the accuracy of our experiment, which we attribute to decoupling of electrons from phonons at these temperatures, so that, below this point, the small residual heat load heats the sample until the heat can be removed by the phonons. We therefore use
\begin{equation}
 G(V_{\rm sd}=0,T)=A \left(T_0^3+T^3\right)^{\frac{\alpha}{3}},\label{eq:G_saturation_tail}
\end{equation}
instead of Eq.~(\ref{eq:G_finite_T}) to fit the lower-temperature exponents, $\alpha_{\rm 2}$ for $V_{\rm g}>-670$~mV and $\alpha_{\rm 1}$ for $V_{\rm g}=-670$~mV. The higher-temperature exponent $\alpha_{\rm 1}$ for $V_{\rm g}>-670$~mV starts at already high enough temperatures that we can ignore the low-temperature saturation and we use Eq.~(\ref{eq:G_finite_T}) to fit it, see the dashed lines in Fig.~\ref{fig:conductance_exponents}A.

\subsection*{Magnetic-field dependence}
\begin{figure}
	\centering
	\includegraphics[width=0.65\linewidth]{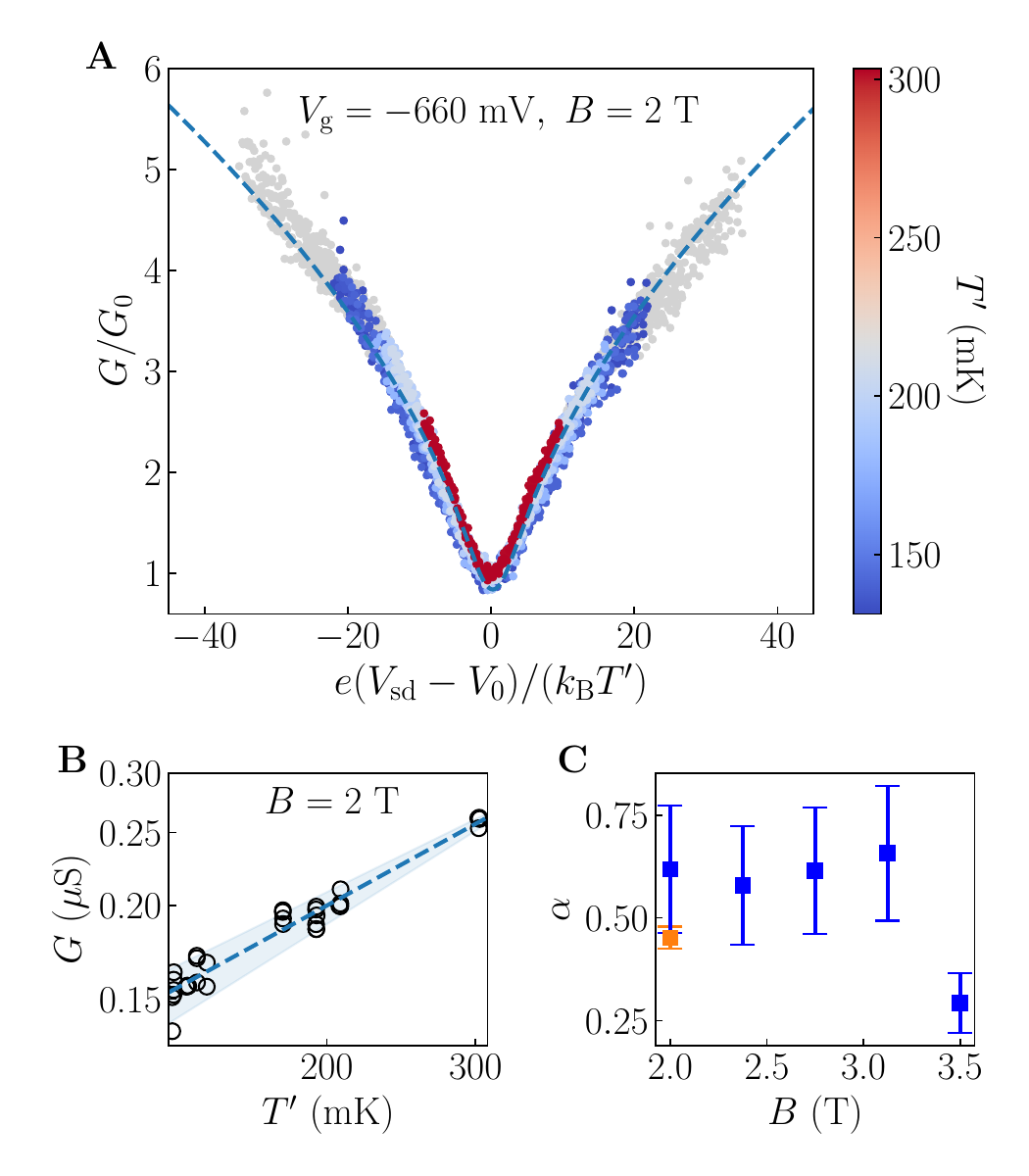}
	\caption{ Magnetic-field dependence of the transport exponent. {\bf A} Rescaled conductance, $G(eV_{\rm sd}/k_{\rm B}T')/G_0$, between 130 and 310\,mK, where $T'$ is the effective electronic temperature allowing for electron heating, for $T_0=130$\,mK, for measurements in a cryostat with more noise heating. The points are colored according to the temperature scale shown on the color bar on the right. The gray points are outside of the linear regime, $|V_{\rm sd}|>0.25$~mV, and are excluded from the fit.
		{\bf B} Conductance at $V_{\rm sd}=0$ as a function of temperature on a log-log plot. The dashed blue line is a fit to Eq.~(\ref{eq:G_finite_T}) with $\alpha=0.58$, which has a relatively large uncertainty in the parasitic background conductance of about $\pm 0.02$\,$\mu$S, shown by the light blue shading. The data in {\rm A} and {\rm B} were measured at $B=2$~T. {\bf C} The blue points show the $B$-field dependence of $\alpha$ with large error bars derived from those in {\rm B} and the uncertainty in the relatively large $T_0$. The orange point is the interpolated value of $\alpha_1$ from Fig.~\ref{fig:conductance_exponents}B for the lower-temperature experimental run. All these data were measured in the single-subband regime at $V_{\rm g}=-660$\,mV, for which $B^+=3.13$\,T.}
	\label{fig:magnetic_field_dependence}
\end{figure}

The magnetic-field dependence of the tunneling exponents was investigated separately, in a different dilution refrigerator with a base temperature below 60\,mK, but with less noise filtering and hence higher electron heating. Fig.~\ref{fig:magnetic_field_dependence}A shows the rescaled conductance $G(eV_{\rm sd}/k_{\rm B}T^\prime)/G_0$ as in Fig.~\ref{fig:conductance_1subband}E for $B=2$\,T, from which we deduce a minimum electron temperature of $T_0= 130$\,mK. From similar plots and fits for different magnetic fields, the $B$ dependence of $\alpha$ is determined (see Fig.~\ref{fig:magnetic_field_dependence}C). The transport Luttinger-liquid exponent $\alpha$ remains largely momentum-independent within the field range $B^-$ to $B^+$ ($B^-=0.70$\,T, $B^+=3.13$\,T for the value of $V_{\rm g}$ in this figure),\textit{ i.e.}, between the $\pm k_{\rm F}$ points, as expected for the Tomonaga-Luttinger theory \cite{Giamarchi_book}. 

However, there appears to be a significant reduction of the exponent $\alpha$ for $B>B^+$, \textit{i.e.}, for $k>k_{\rm F}$. We have previously observed signatures of this behavior in the exponent of the voltage dependence in \cite{Jin19}. Such a reduction could be a hint of the spin-charge separation of the whole Fermi sea beyond the linear regime \cite{Vianez21}. The emerging theory of nonlinear Luttinger liquids has already predicted a second linear Luttinger liquid around the $3k_{\rm F}$ point as a result of the spin-charge splitting of the Fermi surface \cite{Tsyplyatyev22}, with the second Luttinger liquid consisting of only the charge (density-wave) modes. On the qualitative level, this prediction implies a reduction of the transport exponent calculated in Eq.~(\ref{eq:alpha_bulk}) since only the charge modes (with the same Luttinger parameters as around the $k_{\rm F}$ point) contribute to it under the sum over $\nu$, which is in agreement with our observation in Fig.~\ref{fig:magnetic_field_dependence}C. We stress here that a transport theory still needs to be developed to make a quantitative interpretation of such an effect in our data.

\subsection*{Spectroscopy}
Before we proceed to interpretation of the measured transport exponents, we extract another piece of information from our data. In the nonlinear regime away from the $V_{\rm sd}=0$ line, the spin- and charge-density-wave modes fill their corresponding Fermi seas \cite{Vianez21,Tsyplyatyev22}, manifesting themselves as two parabolic dispersions with different masses, which we also observe in our data---see the green and pink dashed lines in Fig.~\ref{fig:conductance_1subband}A. Close to the Fermi points $\pm k_{\rm F}$, these pairs of dispersive lines converge in the linear low-energy region of $V_{\rm sd}=0$, allowing us to extract the two microscopic parameters of the linear Luttinger liquid, the renormalized velocity $v_\nu$ and the dimensionless Luttinger parameter $K_\nu$ directly. Here, the spin-charge separation effect doubles the number of these parameters due to lifting of the degeneracy between the charge ($\nu={\rm c}$) and spin ($\nu={\rm s}$) degrees of freedom.

Focusing our analysis around the $+k_{\rm F}$ Fermi point now, we fit two slopes in our data, see the two black lines converging on the $B^+$ point in Fig.~\ref{fig:conductance_1subband}B as an example. The spin line produces a maximum in $G$, which is clearly visible as a white line in the hole sector ($V_{\rm sd}<0$) in the $B$-derivative in Fig.~\ref{fig:conductance_1subband}B and in the $V_{\rm sd}$-derivative in Fig.~\ref{fig:conductance_1subband}A. The charge line, on the other hand, represents a drop in conductance, where many-body excitations cease to be possible, and, being steeper, shows as a clear minimum only in the $B$-derivative in the hole sector, which makes it less visible \cite{Altland99}. However, it still produces a maximum in $G$ in the particle sector, which has a good visibility as a white line in the $V_{\rm sd}$-derivative in our experiment. From the slopes, we extract the two gradients $\Delta E_{\nu}/\Delta B$. They are converted to a pair of velocities as $v_{\nu}=\Delta E_{\nu}/\left({ed\Delta B}\right)$ using the momentum shift $edB$ in the electron tunneling between two wells, see details in Methods, and the center-to-center separation between the wavefunctions in each well $d=32$\,nm obtained from the band-structure calculation for the design of our double-well heterostructure, see details in \cite{Vianez23}.
The velocities obtained in this way for the whole range of $V_{\rm g}$ that we used are presented in Fig.~\ref{fig:conductance_exponents}C. The error bars there are reduced due to stability of the spin and charge modes in the whole band, so the fitting of two parabolas improves the accuracy of extracting their slopes at the Fermi points. The data points on this figure were always extracted for the first, highest-density 1D subband.

Simultaneously, we measure the distance between the two points ($B^+-B^-$) at which the 1D dispersion crosses the $V_{\rm sd}=0$ line (see, e.g., Fig.~\ref{fig:conductance_1subband}A). This difference gives the Fermi velocity of the 1D system as $v_{\rm F}=ed\left(B^+-B^-\right)/\left(2 m_0\right)$, where we use the value of the single-particle electron mass in GaAs, $m_0=0.0525 m_e$, that was recently measured in \cite{Vianez23}. The Fermi velocities for the first, highest-density 1D subband for all measured values of $V_{\rm g}$ are presented as black squares in Fig.~\ref{fig:conductance_exponents}C. They increase as $V_{\rm g}$ becomes less negative, since that increases the 1D electron density $n_{\rm 1D}=2v_{\rm F} m_0/(\pi \hbar)$.

Together with the pairs of values of $v_{\rm c}$ and $v_{\rm s}$, this information is sufficient to extract the other dimensionless Luttinger parameters for a Galilean-invariant system as $K_\nu=v_{\rm F}/v_\nu$ \cite{Haldane81}. The obtained values of these dimensionless Luttinger parameters are presented in Fig.~\ref{fig:conductance_exponents}D. For more positive $V_{\rm g}$, $n_{\rm 1D}$ increases, so the interaction parameter $r_{\rm s}=1/\left(2a_{\rm B}'n_{\rm 1D}\right)$ decreases, where $a_{\rm B}'$ is the Bohr radius of conduction electrons in GaAs. 
Therefore, as $V_{\rm g}$ becomes more positive, weaker interactions make the difference between the dimensionless Luttinger parameters $K_\nu$ smaller, tending towards their non-interacting limit $K_{\rm c}=K_{\rm s}=1$, in which $v_{\rm c}$ and $v_{\rm s}$ become the same and equal to $v_{\rm F}$ for free fermions \cite{Giamarchi_book}.
  
\section*{Discussion}
We now interpret the transport data quantitatively, and start from the conductance measured at zero $V_{\rm sd}$ in Fig.~\ref{fig:conductance_exponents}A. The low-temperature part of these data is in the linear regime, where the Tomonaga-Luttinger model is applicable. The extent of this region can be estimated from the voltage that separates the linear from the nonlinear energy regions in the single-subband regime in Fig.~\ref{fig:conductance_1subband}A, $V_{\rm sd}=0.25$\ mV, as $T= 0.25{\rm mV}\cdot e/\left(3 k_{\rm B}\right)\simeq 1$~K, where the numerical factor of $3$ between $V_{\rm sd}$ and $T$ was established phenomenologically in the experiment on semiconductor wires in \cite{Jompol09}. We ignore data above a slightly higher temperature $T>2-3$~K in Fig.~\ref{fig:conductance_exponents}A since the chemical potential is larger at higher densities, extending the linear regime to somewhat higher values of $V_{\rm sd}$. 

\begin{figure}[t!]
	\centering
	\includegraphics[width=\linewidth]{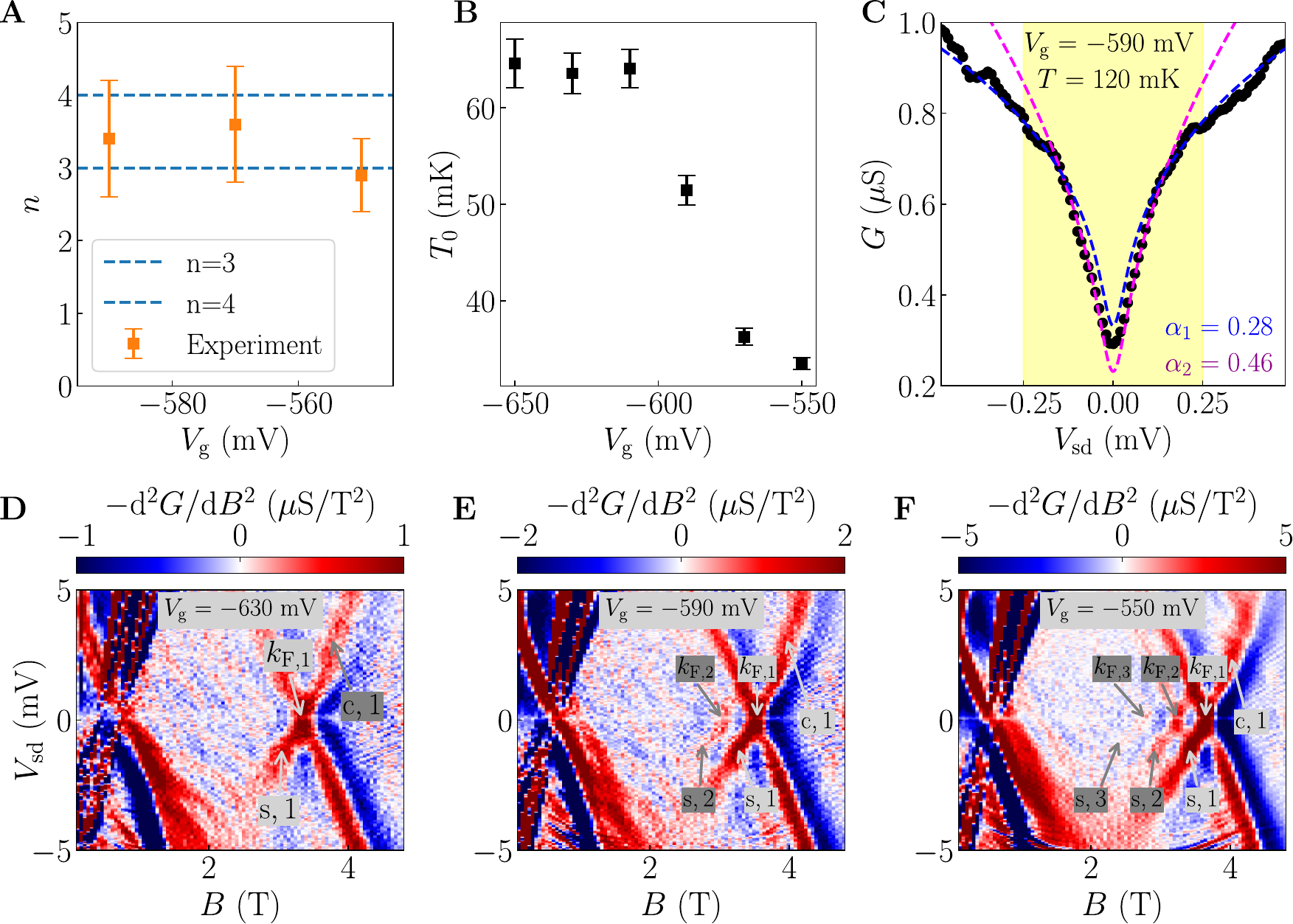}
	\caption{Saturation of conductance at low temperatures and occupation of higher subbands. 
		{\bf A} Saturation exponents $n$ obtained from fitting the low-temperature data in Fig.~\ref{fig:conductance_exponents}A to $G\propto\left(T_0^n+T^n\right)^{\frac{\alpha}{n}}$ for $V_{\rm g}=-550,-570$ and $-590$\,mV with the error bars indicating the rms error in the fit. {\bf B} Saturation temperatures $T_{\rm 0}$ obtained in the same fit for the full range of $V_{\rm g}$ with the error bars indicating the range of values that give an acceptable fit.
		{\bf C} Voltage cut for  $V_{\rm g}=-590$\ mV with two occupied subbands and low temperature, $T=120$\,mK. The dashed lines are Eq.\ (\ref{eq:G_finite_T}) with two exponents $\alpha_{\rm 1}=0.28$ (blue line) and $\alpha_{\rm 2}=0.46$ (magenta line) obtained by fitting the corresponding regions in the data in this voltage cut. The crossover voltage between the two exponents is $V_{\rm sd}=0.12$\ mV.
		{\bf D-F} Evolution of G($B$, $V_{\rm sd}$) as the finger-gate voltage is decreased, for $V_{\rm g}=-630,-590$ and $-550$\,mV. The negative of the second-order derivative of the conductance $G$ with respect to the magnetic field $B$ is plotted, in which the maximum of the signal corresponds to the centers of the lines. From {\bf D} to {\bf F}, more subbands are populated, as can be seen by the appearance of additional crossings around $k_{\rm F,(1,2,3)}$. The labels $(\rm c,s),(1,2,3)$ mark the nonlinear spinon and holon modes away from the linear region, which form the Fermi points for each subband where they cross the Fermi level.
	}
	\label{fig:fig_lowT_subbands}
\end{figure}

At $V_{\rm g}=-670$\,mV, which corresponds to the lowest electron density of $n_{1D}= 37\,\mu\rm m ^{-1} $ in the wires that we measure, only the lowest 1D subband is occupied and we observe only a single slope in conductance, corresponding to a single power law with the exponent $\alpha_1$ going for well over a decade from $T=1$\,K down to about $60$\,mK on the log-log scale in Fig.~\ref{fig:conductance_exponents}A. Below $T\simeq 60$\,mK the conductance saturates at a constant value that originates most likely from thermal coupling bottlenecks common at millikelvin temperatures, making even small parasitic heat sources balance out the limited cooling power and keeping the electronic temperature above that of the cryostat. 
In order to do a quantitative assessment in this regime, we construct phenomenologically the formula  $G\sim\left(T_0^n+T^n\right)^{\alpha/n}$, which describes interpolation between the Luttinger-liquid power law $G\sim T^\alpha$ at $T\gg T_0$ and a saturation tail $G-G(T=0)\sim T^n$ at $T\ll T_0$. Using $n$ and $T_{\rm 0}$ as fitting parameters, we find their values in Fig.~\ref{fig:fig_lowT_subbands}A and B. The low signal-to-noise ratio prevents us from performing this analysis in the single-subband regime. However, as we make $V_{\rm g}$ less negative, the current and thereby the signal-to-noise ratio increase, allowing us to see the shape of the bending from the power law to the constant for $V_{\rm g}\geq-590$\,mV.

The statistical error for $n$ in Fig.~\ref{fig:fig_lowT_subbands}A is smallest for the highest density, since the low-temperature conductance becomes large enough to see the onset of saturation move to well below $60$~mK, giving more reliably $n=3$ for $V_{\rm g}=-550$~mV, but the amplitude of the signal decreases rapidly with decreasing density, giving a less-defined $n=3$ or $4$ for $V_{\rm g}=-570$ and $-590$~mV. Altogether, the current data, given the current state of the art, do not select a particular exponent for the saturation tail but rather restrict it to the range $n=3-4$. These exponents are close to but systematically smaller than the $n=5$ prediction of the purely electron-phonon mechanism in 3D bulk \cite{Wellstood94}, which suggests an additional cooling process such as out-diffusion of electrons, i.e., Wiedemann-Franz cooling \cite{Meschke04,Palma17}. 
The fitted values of $T_{\rm 0}$ in Fig.~\ref{fig:fig_lowT_subbands}B are well-defined for all $V_{\rm g}$, showing a two-fold decrease when the second subband is occupied, which could indicate additional cooling due to the Wiedemann-Franz process since the higher electronic density in the wires also increases the conductance through the whole structure somewhat. For the sake of concreteness, we use $n=3$ in the formula for conductance in Eq.~(\ref{eq:G_saturation_tail}) and for the electronic temperature in $T^\prime$ that we used to fit the Luttinger-liquid exponents in Fig.~\ref{fig:conductance_exponents}A. 

Continuing the analysis of the zero-voltage conductance in Fig.~\ref{fig:conductance_exponents}A, we consider the whole temperature range for  $V_{\rm g}>-670$\,mV. In this gate-voltage range a second exponent $\alpha_2$ appears below an intermediate temperature of about $400$\,mK, and both exponents $\alpha_1, \alpha_2$ evolve with $V_{\rm g}$, see the blue and magenta squares in Fig.~\ref{fig:conductance_exponents}B. The main physical process behind these power laws can be assessed by comparing the directly measured transport exponent with the predictions of the Tomonaga-Luttinger theory. One of the two possibilities is electrons tunneling at any point along the quantum wire \cite{Kane97,Altland99}, which is known as bulk tunneling and is expected to dominate in infinitely long wires. The other is local tunneling at an end of the finite wire \cite{Kane92,Tserkovnyak03}, which is usually referred to as end tunneling and has Friedel oscillations mixed in on top of the Luttinger-liquid density modes \cite{Fabrizio95,Eggert96}. The conductance in both regimes is evaluated within the framework of the Tomonaga-Luttinger model relating the transport exponent to the microscopic Luttinger liquid parameters as
\begin{align}
\alpha_{\rm bulk}&=\sum_{\nu=\rm s,c}\frac{K_{\nu}+K_{\nu}^{-1}-2}{4},\label{eq:alpha_bulk}\\
\alpha_{\rm end}&=\frac{K_{\rm c}^{-1}+K_{\rm s}^{-1}-1}{2},\label{eq:alpha_end}
\end{align}
see details in Supplementary Note 1. The results of our comparison are plotted as the black and green squares, respectively, in Fig.~\ref{fig:conductance_exponents}B. The microscopic parameters $v_{\nu}$ and $K_{\nu}$ for the Tomonaga-Luttinger model are readily measured as a function of $V_{\rm g}$ using transport spectroscopy in the same sample as used in Fig.~\ref{fig:conductance_exponents}C and E. Since both $\alpha_1$ and $\alpha_2$ are about an order of magnitude larger than the predicted value of $\alpha_{\rm bulk}$ and are of the same order as $\alpha_{\rm end}$, we conclude that both transport exponents originate mainly from the end-tunneling process. 

Following this conclusion, we attribute the appearance of the second exponent to occupation of the second 1D subband in the quantum wire. A simple model describing the conductance measured in our experiment at low energy can be constructed by treating two subbands as a pair of conductors connected in parallel. The electrons can enter either of the two subbands from the same 2DEG in the upper well and tunnel from either of the subbands to the 2DEG in the bottom well independently, see the sketch in Fig.~\ref{fig:device}. The total conductance, then, is the sum of two individual conductances,
\begin{equation}
G=A_{1}\min\left(T,T_1\right)^{\alpha_1'}+A_{2}\min\left(T,T_2\right)^{\alpha_2'},\label{eq:conductance_parallel}
\end{equation}
where the parameters $A_i$, $\alpha_i'$, and $T_i$ are different for each of the two subbands. Since $\alpha_1<\alpha_2$ for each gate voltage in Fig.~\ref{fig:conductance_exponents}B, $\alpha_2'$ has to be attributed to the second subband, which has a smaller density and therefore larger $r_{\rm s}$, leading to stronger interaction effects. The min functions in this equation embody the applicability limit of the linear Tomonaga-Luttinger theory. Beyond the energy $k_{\rm B}T_i$, the power-law increase of the conductance ceases and we model (very) crudely the transport for the nonlinear theory at small momenta corresponding to $B=2$\,T as a constant, motivated by our observation in the voltage cuts in Fig.~\ref{fig:conductance_1subband}E, that the gray points in the nonlinear region lie systematically below the blue dashed power-law curve. We have already estimated $T_1\simeq 1$\,K for the first subband. For the second subband, $T_2\simeq 400$\,mK is somewhat smaller, owing to the lower density, which results in a smaller chemical potential and therefore in a smaller extent of the linear region.

The whole dataset in Fig.~\ref{fig:conductance_exponents}A can be explained with these values of $T_i$,  a pair of amplitudes $A_1<A_2$, and a pair of $\alpha_1'>\alpha_1$, $\alpha_{\rm 2}'>\alpha_{\rm 2}$, in which the latter is due to the total conductance in Eq.~(\ref{eq:conductance_parallel}) always being a sum of two contributions. At low temperatures $T<T_{\rm 1},T_{\rm 2}$, the second contribution, with the larger exponent $\alpha_{\rm 2}'$, dominates, but the smaller exponent $\alpha_{\rm 1}'$ reduces the effective value $\alpha_{\rm 2}$ in $G$ to $\alpha_{\rm 1}'<\alpha_{\rm 2}<\alpha_{\rm 2}'$. At high temperatures  $T_{\rm 2}<T<T_{\rm 1}$, the first contribution with the smaller exponent $\alpha_{\rm 1}$ dominates in  Eq.~(\ref{eq:conductance_parallel}) but the second contribution is still a constant, acting as the exponent $\alpha_{\rm 2}=0$, and reducing $\alpha_{\rm 1}$ in $G$ to $\alpha_{\rm 1}<\alpha_{\rm 1}'$. Note that the bulk-tunneling process is always present in our experiment since the electrons can tunnel from any position in the wire to the 2DEG in the bottom well through the same tunneling barrier. This process occurs in parallel with the end-tunneling process, so we always need to add its contribution $A_{\rm bulk} T^{\alpha_{\rm bulk}}$ to the conductance in Eq.~(\ref{eq:conductance_parallel}). However, since $\alpha_{1,2}\gg\alpha_{\rm bulk}$ the contribution from the bulk-tunneling process (with much smaller exponent) is much smaller for large enough $T$. We were unable to observe it independently down to the smallest $T_0\simeq 35$\,mK seen in our experiment, although it is possible that it explains some or all of the saturation itself. 

By measuring a voltage cut ($G$ as a function of $V_{\rm sd}$) at a higher electron density at $V_{\rm g}=-590$\,mV and at an intermediate temperature of $T=120$\,mK above $T_0$ but below $T_1$, we find further evidence for the two-subband interpretation. Fitting the data in Fig.~\ref{fig:fig_lowT_subbands}C with Eq.~(\ref{eq:G_finite_T}) we find two exponents in the linear regime of $\left|V_{\rm sd}\right|<0.25$\,mV: $\alpha_2=0.46$ at smaller $V_{\rm sd}$ and $\alpha_1=0.28$ for larger $V_{\rm sd}$. Within the relatively large uncertainty of this fit (of about $20$\%) these two exponents are the same exponents $\alpha_1$ and $\alpha_2$ in Fig.~\ref{fig:conductance_exponents}B for $V_{\rm g}=-590$\,mV measured in $G$ at $V_{\rm sd}=0$ as a function of $T$. The crossover point in voltage at $V_{\rm sd}=0.12$\,mV gives the same crossover temperature (within error bars) of $T_{\rm 2}=0.12\,\textrm{mV}\cdot e/\left(3k_{\rm B}\right)\simeq 450$\,mK that we observe in the temperature-resolved measurements of $G$ at $V_{\rm sd}=0$ in Fig.~\ref{fig:conductance_exponents}A.

In the spectroscopic maps that we measure as $G$ in a wide range of $V_{\rm sd}$ and $B$ covering the whole energy band for the same densities corresponding to $V_{\rm g}=-630,-590,-550$\,mV, the second (and third) subband also appears in the form of the second (and third) pair of the spin charge parabolae, see Fig.~\ref{fig:fig_lowT_subbands}D-F. In this figure, the second (and third) sets of parabolae marked by $(\rm s,c),(1,2,3)$ define the second (and third) Fermi points marked by $k_{\rm F,(1,2,3)}$ that correspond to successively smaller densities of the higher 1D subbands in our quantum wires. While the appearance of the second transport exponent in the temperature-resolved measurements in Fig.~\ref{fig:conductance_exponents}A generally correlates with the appearance of the second subband in Fig.~\ref{fig:fig_lowT_subbands}D-F, the second subband in Fig.~\ref{fig:fig_lowT_subbands}D-F appears at somewhat higher $V_{\rm g}$ than the second exponent. This happens since the ZBA hinders the low-energy sector up to a finite value of $V_{\rm sd}$ in the transport spectroscopy measurements, e.g., up to $V_{\rm sd}=0.25$\,mV in Fig.~\ref{fig:conductance_1subband}A. In order for the second subband to be visible, the density has to become large enough for its chemical potential to exceed this threshold. For the lowest $V_{\rm g}=-550$~mV that we investigated, the crossover region in the transport exponent in Fig.~\ref{fig:conductance_exponents}A around $T=400$\,mK widens, which hints at a third exponent developing in between $\alpha_1$ and $\alpha_2$, corresponding to the appearance of the third subband in Fig.~\ref{fig:fig_lowT_subbands}F. However, the extent of this region in Fig.~\ref{fig:conductance_exponents}A is still too small (narrower than a decade in temperature) to draw a definitive conclusion.

\section*{Methods}
\subsection*{Sample preparation}
All out-of-wire tunneling devices measured in this work were fabricated using GaAs/AlGaAs heterostructures grown via molecular-beam epitaxy (MBE), and composed of two identical $18$\,nm quantum wells (QWs) separated by a $14$\,nm-thick GaAs/AlGaAs superlattice barrier. Si-doped layers on the far side of each well lead to electron densities of $2.85 (1.54)\times10^{15}$\,m$^{-2}$ and mobilities of $191 (55)$\,m$^{2}$V$^{-1}$s$^{-1}$ in the top (bottom) wells, as measured by the Shubnikov--de-Haas effect at $1.4$\,K. 

Ti/Au gates were patterned using a combination of photo- and electron-beam lithography, see Fig.\ \ref{fig:device}. Electrical contact to both wells was achieved via standard AuGeNi ohmic contacts. Gates were then biased to inject current from one ohmic contact through the 1D channel defined only in the upper well by the split gates and mid-gate. The current was then carried by electrons tunneling to or from the lower well in the central array of 1D wires, and it then flowed out beneath the barrier gate (which blocked the upper well) to the other ohmic contact (see \cite{Vianez23} for further details).

Our spectroscopy technique allows us to probe the dispersion of a given system (\textit{e.g.}, a 1D array of wires) with respect to a known standard (\textit{e.g.}, a 2D Fermi liquid) by measuring the tunnel current between both. This is given by the convolution of the two spectral functions as \cite{Altland99}
\begin{multline}
  I\left(B,V_{\rm sd}\right)=\int {\rm d}^2 {\bf k} {\rm d}\varepsilon \left(f^{\rm UW}_T(\varepsilon-eV_{\rm sd})-f^{\rm LW}_T(\varepsilon)\right)\\
  A_{\rm UW}\left({\bf k},\varepsilon\right)A_{\rm LW}\left({\bf k}+ed\left({\bf n}\times{\bf B}\right)/\hbar,\varepsilon-eV_{\rm sd}\right),\label{eq:current}
\end{multline}
where $A_{\rm UW/LW}\left({\bf k},\varepsilon\right)$ and $f^{\rm UW/LW}_T(\varepsilon)$ are the spectral functions and the Fermi distribution of the electrons in the upper/lower wells (UW/LW), $-e$ is the electron charge, $d$ is the distance between the wells, ${\bf n}={\bf\hat{z}}$ is the normal to the 2D plane. In order to map the full dispersion of each system, we then measure the differential conductance $G=\textrm{d}I/\textrm{d}V$ as a function of both energy $\varepsilon$ and momentum $\hbar\bf k$. This is achieved by simultaneously applying a DC bias $eV_\textrm{sd}$ between the layers (\textit{i.e.}, offsetting their Fermi energies) and varying the in-plane magnetic field $B$ applied in the direction perpendicular to the wires ${\bf B}=-B{\bf\hat{y}}$, so that the momentum of the tunneling electrons is shifted by $edB$ in the $x$-direction.

\subsection*{Conductance measurements}
In this work, we measure the differential conductance between the two wells, $G\left(B,V_{\rm sd}\right)=\partial_{V_{\rm sd}} I\left(B,V_{\rm sd}\right)$. In order to achieve low electron temperatures, the measurement lines were filtered by a two-stage $RC$ low-pass filter and subsequently passed through inductive microwave filters. $G$ was measured using a lock-in amplifier at low frequency ($17.77$\,Hz) with a small ac excitation of 2--6\,$\mu$V rms. The line resistance was calibrated on the first conductance plateau of the split-gate characteristic, and subsequently subtracted.

When the wires are completely pinched off ($V_{\rm g}<-700$\,mV), the transport is purely in the 2D--2D tunneling regime, since there is still a non-negligible `parasitic' area of 2DEG that takes current from the injector to the 1D wires, see Fig.\ \ref{fig:device}. The current in this regime is described by the 2D Fermi liquid in both wells. Its spectral functions $A_{\rm UW/LW}({\bf k},\varepsilon)=\delta(\varepsilon-\varepsilon_{\rm 2D}({\bf k}))$ are centered on parabolae 
\begin{equation}
  \varepsilon_{\rm 2D}({\bf k})=\frac{\hbar^2 \left(k-k_{\rm F, L/U}^{\rm 2D}\right)^2}{2m^*_{\rm 2D}}, \label{eq:eps2D}
\end{equation}
with the effective mass $m^*_{\rm 2D}$ renormalised by the Coulomb interaction according to the Landau's Fermi-liquid theory; the Fermi wave-vectors are $k_{\rm F,U}^{\rm 2D}$ and $k_{\rm F,L}^{\rm 2D}$, respectively. Substitution of these spectral functions in Eq.\ (\ref{eq:current}) models two parabolic dispersions in the conductance. The peaks in our data are fitted well by this model with $d=32$\,nm and $m^*_{\rm 2D}=0.062\,m_{\rm e}$, where $m_{\rm e}$ is the free-electron mass, in the same way as it was in \cite{Vianez21}.

When reducing $V_{\rm g}$, the tunnel current in our device has two contributions. One is from the tunneling through the array of 1D wires to the lower 2DEG (which we are interested in) and the other is from the tunneling through the 2D `p' region. This parasitic tunneling leads to uncertainties in the extraction of the tunneling exponents and, therefore, has to be accounted for. To do so, we measure the conductance as a function of $V_{\rm g}$ past wire pinch-off and observe that the remaining 2D--2D conductance is linear in $V_{\rm g}$. We therefore extrapolate the linear dependence to the $V_{\rm g}$ of interest and subtract it from the measured conductance. Such subtraction of the parasitic 2D--2D signal is performed in all measurements of the wires, taking the uncertainties into account in the overall error estimates.

\subsection*{Low-temperature setup}
Except where noted, all measurements were carried out in a heavily modified wet dilution refrigerator that is optimized for achieving ultra-low temperatures \cite{Casparis12}. Each lead is connected through a thermocoax running down to the mixing chamber, which acts as an excellent microwave filter for frequencies above 3\,GHz. The leads are then thermally anchored to the mixing chamber using silver-epoxy microwave filters \cite{Scheller14} offering $>100$\,dB attenuation above $200$\,MHz. A 2\nobreakdash-pole discrete component $RC$-filter board reduces the final bandwidth down to a few kHz. Subsequently, each measurement wire runs through the mixing chamber, where sintered-silver heat exchangers, each with an effective surface area of 3\,m$^2$, guarantee optimal lead thermalization down to the lowest temperatures, thus allowing efficient electronic Wiedemann-Franz cooling through the measurement leads on low-impedance devices. For resistive devices, on the other hand, thermalization occurs predominantly by phonon cooling through the sample substrate. Electronic sample temperatures down to 10\,mK have been measured using quantum-dot thermometry in a GaAs 2DEG \cite{Maradan14}. The present device, mounted on a Kyocera leadless chip carrier with heat-sunk gold backplane, is resistive enough that the latter process should dominate.

\section*{Data availability} The Basel data generated in this study are available at Zenodo\linebreak (http://doi.org/10.5281/zenodo.15639288) and all the data are available at the University of Cambridge data repository (http://doi.org/10.17863/CAM.119078).

\printbibliography

\section*{Acknowledgments} C.J.B.F and P.M.T.V. acknowledge funding from the UK EPSRC (Grant No. EP/J01690X/1 and EP/J016888/1), an EPSRC International Doctoral Scholars studentship (Grant No.\linebreak EP/N509620/1) and an EPSRC Doctoral Prize, as well as support from the Horizon 2020 European Microkelvin Platform. D.M.Z., C.P.S., H.W., and O.S.S. acknowledge support from the NCCR SPIN and Grant No. 215757 of the Swiss NSF, the Georg H. Endress Foundation, the EU H2020 European Microkelvin Platform EMP (Grant No. 824109) and FET TOPSQUAD (Grant No. 862046). O.T. acknowledges funding from the DFG (Project No. 461313466).

\section*{Author Contributions} Y.J., M.M., and W.K.T. fabricated the experimental device, with H.W., O.S.S., C.P.S., and P.M.T.V. performing the transport measurements shown, except for the magnetic-field dependence, which was done by Y.J., M.M., W.K.T. and C.J.B.F.. J.P.G. performed the electron-beam lithography and I.F. and D.A.R. grew the heterostructure material.
H.W., C.J.B.F. and O.T. analyzed the data. O.T. developed the theoretical framework. C.J.B.F. and D.M.Z. supervised the experimental side of the project. All authors contributed to the discussion of the results. O.T. and C.J.B.F. wrote the manuscript.

\section*{Competing interests} The authors declare that they have no competing interests.

\section*{Corresponding authors}
Correspondence should be addressed to Oleksandr Tsyplyatyev (tsyplyatyev@itp.uni-frankfurt.de), Christopher J.B. Ford (cjbf@cam.ac.uk), and Dominik Zumb\"uhl (dominik.zumbuhl@unibas.ch).

\end{document}

\maketitle
\newpage

\section{Supplementary Note 1: Tomonaga-Luttinger theory}
%Having measured  separately the transport exponent $\alpha$ in Fig.~\ref{fig:fig_exponent_luttinger_parameters}(Top) and the microscopic parameters $K_\nu$, $v_\nu$ in Fig.~\ref{fig:fig_exponent_luttinger_parameters}(Bottom) of the interacting one-dimensional electronic system, we can check consistency of the two experimental results by means of the Tomonaga-Luttinger theory \cite{Giamarchi_book} that predicts a particular relation between these two quantities. 
Here, we briefly summarise the Tomonaga-Luttinger theory \cite{Giamarchi_book} that predicts a particular relation between the transport exponent $\alpha$ and  the microscopic parameters $K_\nu$, $v_\nu$.

The Tomonaga-Luttinger model, describing interacting one-dimensional electrons with spin-1/2 after the bosonisation in the low-energy regime, is given by the following Hamiltonian \cite{Tomonaga50,Luttinger63},
\begin{equation}
    H=\int {\rm d}x  \sum_{\nu=s,c} \frac{v_\nu}{2\pi}\left[ K_\nu \left(\nabla \theta_\nu(x) \right)^2 +\frac{\left(\nabla \varphi_\nu(x) \right)^2}{K_\nu }\right],\label{eq:Hspinful}
\end{equation}
where $v_\nu$ are the renormalised velocities of the collective modes, $K_\nu$ are the dimensionless  Luttinger parameters describing the interaction strength for the spin (s) and charge (c) degrees of freedom, and the two pairs of the bosonic $\theta_\nu(x)$, $\varphi_\nu(x)$ are canonically conjugated variables, $\left[\varphi_{\nu}(x),\nabla \theta_{\nu'}(x')\right]={\rm i}\pi \delta_{\nu\nu'}\delta\left(x-x'\right)$.

The Green function for the original fermions was evaluated based on this model also using the bosonization technique in \cite{Schoenhammer92,Voit93} as 
\begin{equation}
G^{\pm}\left(x,t\right)= \frac{\pm {\rm e}^{ik_{\rm F}^{\rm 1D}x}}{2\pi\sqrt{x- v_{\rm s}t\pm {\rm i}r}\sqrt{x- v_{\rm c}t\pm {\rm i}r}}
% \times
 \left[\frac{r^{2}}{x^{2}-\left(v_{\rm c}t\mp {\rm i}r\right)^{2}} \right]^{\gamma_{\rm c}}\left[\frac{r^{2}}{x^{2}-\left(v_{\rm s}t\mp {\rm i}r\right)^{2}}\right]^{\gamma_{\rm s}}, \label{eq:spinfulG}
\end{equation}
where $\gamma_\nu=(K_{\nu}+K_{\nu}^{-1}-2)/8$, the $\pm$ sign marks the particle and hole sectors, $k_{\rm F}^{\rm 1D}$ is the Fermi momentum, and $r$ is a small but finite short-range cutoff. This result gives explicitly the complete information about the static, dynamical and spectral properties of the electrons described by the model in Supplementary Eq.~(\ref{eq:Hspinful}).

\subsection{Bulk-tunneling regime}
The electrons can tunnel from the wire in the upper quantum well to the 2DEG in the lower well at any point along the wire (see the scheme of our device in Fig.~\ref{fig:device}). The electric current  that we measure in this perpendicular geometry is given by the tunneling conductance as the convolution of two spectral functions \cite{Altland99}, which we have already quoted in Eq.~(\ref{eq:current}). Taking the limit of zero temperature $T\rightarrow 0$ and substituting $A_{\rm UW}=A^+_{\rm 1D}$ and $A_{\rm LW}=A^-_{\rm 2D}$, for instance for the positive voltages $V_{\rm sd}>0$ for which electrons tunnel from the wire to the 2DEG, we obtain
\begin{equation}
    I(B,V_{\rm sd})=\int {\rm d}k\int_{eV_{\rm sd}}^{0}{\rm d}\varepsilon A^-_{\rm 1D}\left(k,\varepsilon\right) 
    A_{\rm 2D}^{+}\left(k+edB,\varepsilon-eV_{\rm sd}\right).\label{eq:currentT0}
\end{equation}
Here the spectral function of the quantum wire is given by the Fourier transform of the Green function in Supplementary Eq.\ (\ref{eq:spinfulG}) as 
\begin{equation}
 A_{\rm 1D}^{\pm}\left(k,\omega\right)  =\frac{{\rm i}}{2\pi}\int {\rm d}t\,{\rm d}x\,{\rm e}^{{\rm i}\left(\omega t-kx\right)} G^{\pm}\left(x,t\right)\label{A1D}.
\end{equation}
The spectral function of 2DEG has the $\delta$-functional form centered at the single-particle dispersion in Eq.~(\ref{eq:eps2D}), $A^\pm_{\rm 2D}\left({\bf k},\varepsilon\right)=\delta\left(\varepsilon\mp\varepsilon_{\rm 2D}({\bf k})\right)\theta\left(\pm\varepsilon_{\rm 2D}({\bf k})\right)$ where $\theta(x)$ is the step function. Since in the 1D-2D geometry only the spectral function of the 2DEG depends on $k_y$ in Eq.~(\ref{eq:current}) the integral along this direction can be absorbed into it as $A_{\rm 2D}^{\pm}\left(k_{x},\varepsilon\right)=\int dk_y A^\pm_{\rm 2D}\left({\bf k},\varepsilon\right)$. The resulting projected onto the direction of the wire spectral function of the 2DEG is 
\begin{equation}
A_{\rm 2D}^{\pm}\left(k_{x},\varepsilon\right)=\sqrt{\frac{m^*_{\rm 2D}}{2\hbar^{2}}}\frac{\theta\left(\pm\varepsilon-v_{\rm F,L}^{\rm 2D}\hbar\left(k_{x}+k_{\rm F,L}^{\rm 2D}\right)\right)}{\sqrt{\pm\varepsilon-v_{\rm F,L}^{\rm 2D}\hbar\left(k_{x}+k_{\rm F,L}^{\rm 2D}\right)}},
\end{equation}
where $v^{\rm 2D}_{\rm F,L}=\hbar k^{\rm 2D}_{\rm F,L}/m_{\rm 2D}^*$ is the Fermi velocity of  2DEG.

Evaluation of the integrals in Supplementary Eq.~(\ref{eq:currentT0}) for both positive and negative voltages $V_{\rm sd}$ gives the same the current that is independent of the magnetic field $B$ and has a power-law dependence on the voltage $V_{\rm sd}$, $I(B,V_{\rm sd})\sim\left|V_{\rm sd}\right|^{1+\alpha_{\rm bulk}}$, with the exponent given by the dimensionless Luttinger parameters $K_\nu$ as  
\begin{equation}
\alpha_{\rm bulk}=\sum_{\nu=\rm s,c}\frac{K_{\nu}+K_{\nu}^{-1}-2}{4}~.\label{eq:alpha_bulk}
\end{equation}
The conductance can then be found as a derivative, $G(V_{\rm sd})=\partial_{V_{\rm sd}}I(B,V_{\rm sd})$, giving the transport exponent of the Tomonaga-Luttinger model as 
\begin{equation}
G(V_{\rm sd})\sim\left|V_{\rm sd}\right|^{\alpha_{\rm bulk}}.\label{eq:G_V}
\end{equation}
This power-law vanishing of conductance at small voltages is a signature effect of Luttinger liquids. It is a reflection of a more generic property: the density of states $\rho(\varepsilon)$ for the model in Supplementary Eq.~(\ref{eq:Hspinful})  vanishes at the Fermi energy in the same power-law fashion, $\rho(\varepsilon)\sim\left|\varepsilon\right|^{\alpha_{\rm bulk}}$.
%This is the conductance that we measure in our experiment in the low-energy regime by small $V_{\rm sd}$.

\subsection{End-tunneling regime}
\sloppy The relation between the transport exponent and the microscopic Luttinger parameters in Supplementary Eq.~(\ref{eq:alpha_bulk}) was derived under the assumption of an infinitely long wire. When the length is finite, the bound states at the end provide another local channel for tunneling of the collective modes of Luttinger liquid to the 2DEG in the bottom well. Such a local transport process also results in a power-law dependence of the conductance on voltage $V_{\rm sd}$, in the same way as the non-local tunneling in the previous subsection but with a modified exponent \cite{Kane92,Tserkovnyak03}, in which the Friedel oscillations are mixed in on top of the bulk Luttinger exponent \cite{Fabrizio95,Eggert96}.

The application of the hard-wall boundary condition at $x=\pm L/2$, where $L$ is the length of the wire, to the model in Supplementary Eq.~(\ref{eq:Hspinful}) leads to the modification of its eigenmodes near the edges. Such a modification, in turn, makes the Green function in Supplementary Eq.~(\ref{eq:spinfulG}) explicitly dependent on two coordinates $x,x'$ via multiplication by a finite-size factor as \cite{Eggert96}
\begin{equation}
    G^\pm(x,0;x',t)= G^\pm(x_-,t)
%    \times
    \left[\frac{x_{+}^{2}-x_{-}^{2}}{x_{+}^{2}+\left(v_{\rm c}t\right)^2}\right]^{\gamma_{\rm c}}\left[\frac{x_{+}^{2}-x_{-}^{2}}{x_{+}^{2}+\left(v_{\rm s}t\right)^2}\right]^{\gamma_{\rm s}},\label{eq:spinfulGend}
\end{equation}
where $x_-=x-x'$ and $x_+=x+x'+L$ in the right-hand side are the sum and difference of the two separate spatial coordinates of the Green function. Then the conduction is evaluated using the same steps as in the bulk case, in Supplementary  Eqs.~(\ref{eq:currentT0})-(\ref{eq:G_V}). The only difference is the need to integrate the Green function in Supplementary Eq.~(\ref{eq:spinfulGend}) over sum of coordinates $x_+$ in a small region around the end of the wire to select the the localised end-state before inserting it into the Fourier transform over the difference of the spacial variables in Supplementary  Eq.~(\ref{A1D}). The result at the end is the same conductance as in Supplementary Eq.~(\ref{eq:G_V}) but with the exponent 
\begin{equation}
   \alpha_{\rm end}=\frac{K_{\rm c}^{-1}+K_{\rm s}^{-1}-1}{2}.\label{eq:alpha_end}
\end{equation} 

So far, the conductance was derived at $T=0$ in this section. Introduction of a finite temperature smears the power-law dependence on voltage at low voltages, resulting in the additional temperature dependence in Eq.~(\ref{eq:G_finite_T}), in which the exponent $\alpha$ is the same $\alpha_{\rm bulk}$ in Supplementary~Eq.~(\ref{eq:alpha_bulk}) and $\alpha_{\rm end}$ in  Supplementary Eq.~(\ref{eq:alpha_end}) for both tunneling processes as in the $T=0$ case.

\begin{figure}
    		\centering
    		\includegraphics[width=\linewidth]{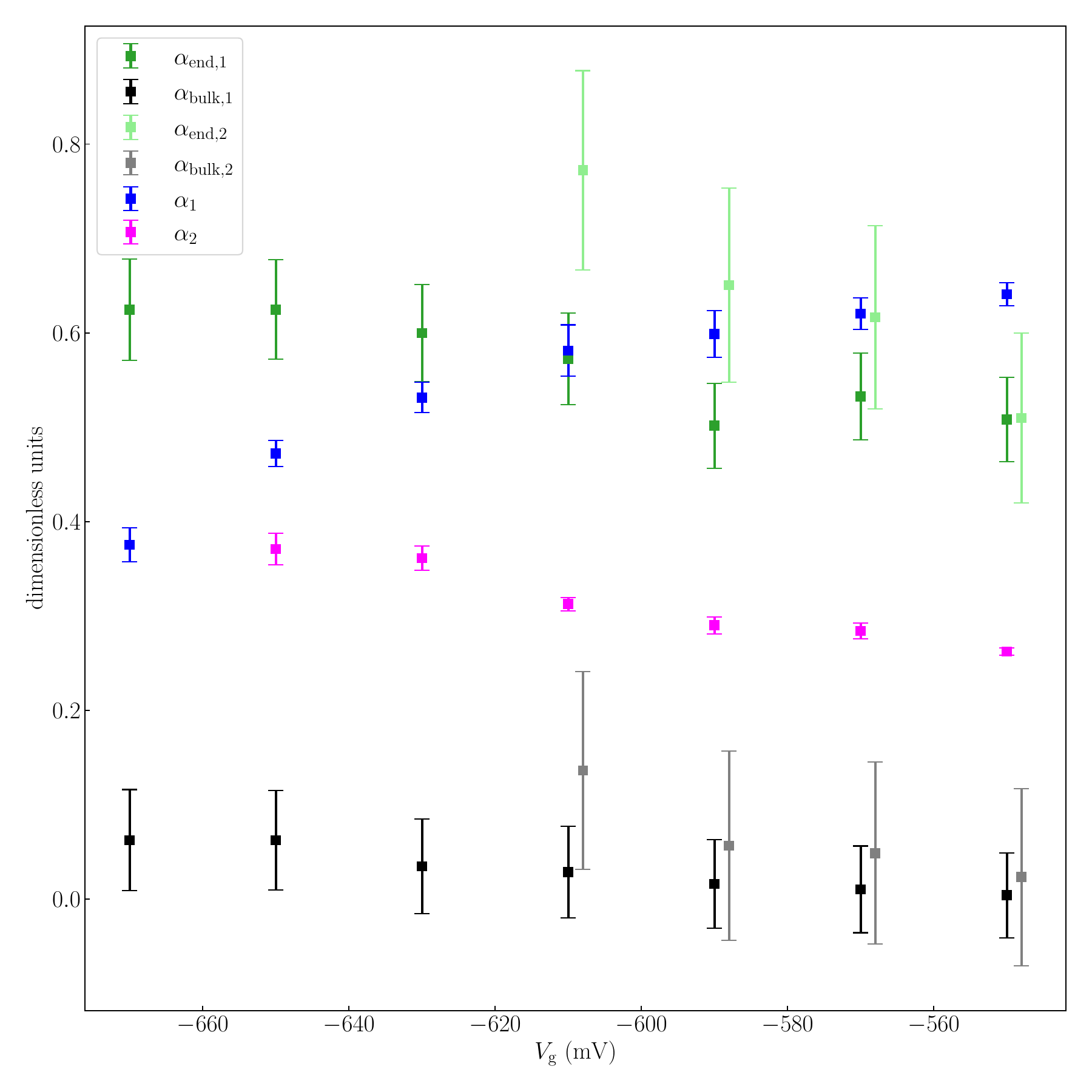}
    		\caption{The values of two exponents $\alpha_{\rm 1}$ (blue squares) and $\alpha_{\rm 2}$ (magenta squares) as a function of $V_{\rm g}$ extracted from the conductance data in Fig.~\ref{fig:conductance_exponents}A with the error bars showing the rms error in the fit. The bulk transport exponent $\alpha_{\rm bulk},i$ (black and gray squares), the end transport exponent $\alpha_{\rm end},i$ (green and the light squares), and their error bars are evaluated for the Luttinger parameters in Fig.\ \ref{fig:conductance_exponents}D using Supplementary Eq.~(\ref{eq:alpha_bulk}) and Supplementary Eq.~(\ref{eq:alpha_end}) respectively. The index $i=1,2$ labels the first and the second subband, when the latter appear at $V_{\rm g}>-620$\,mV. The values for the second subband are estimates only, since they are, in turn, based on the estimates of $K_{\rm s,2}$ and $K_{\rm c,2}$ in Fig.~\ref{fig:velocity}.
      %Extracted tunneling exponents $\alpha$ in the low (green) and high (black) - temperature range. 
      %The bulk ($\alpha_{bulk}$, blue) and the end ($\alpha_{end}$, red) tunneling exponent are calculated using Eq.\ (\ref{eq:alpha_bulk}) and Eq.\ (\ref{eq:alpha_end}) with the Luttinger parameters $K_{c,s}$ from Fig.\ \ref{fig:velocity}.
      }
    		\label{fig:fig_exponent}
        \end{figure}  
        
\begin{figure}
    		\centering
    		\includegraphics[width=0.6\linewidth]{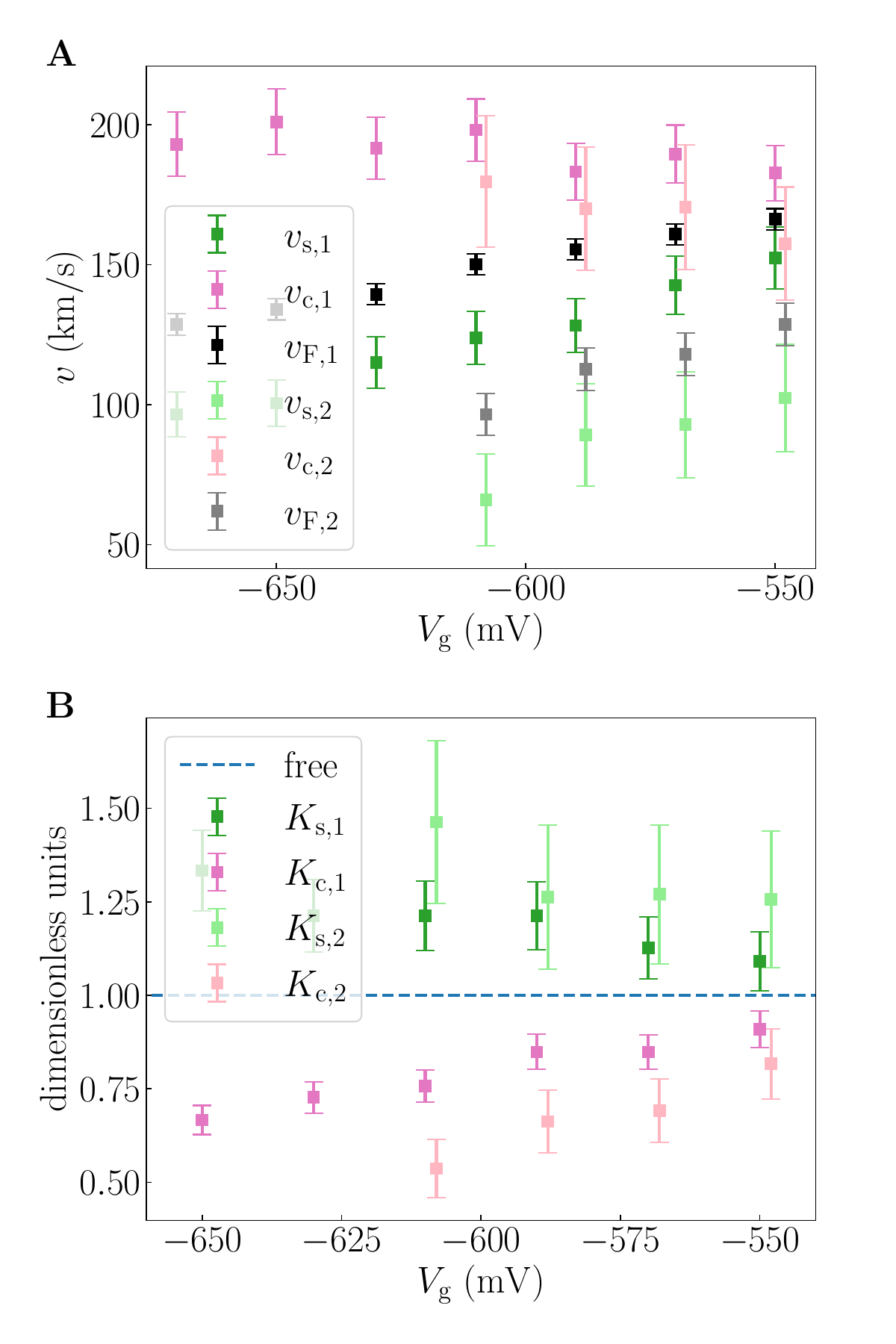}
    		\caption{Estimate of the Luttinger liquid parameters for the second subband. \linebreak {\bf A} The spin ($v_{{\rm s},i}$, green squares) and charge ($v_{{\rm c},i}$, pink squares) velocities extracted from the spectroscopic maps, \emph{e.g.}, Fig.~\ref{fig:conductance_map_1subband}B, as the  linear slopes around the $B_+$ point, the Fermi velocity extracted as the distance between the $B_\pm$ points, and the error bars indicate the range of values that give an acceptable fit. {\bf B} The Luttinger parameters for spin ($K_{{\rm s},i}$, green squares), charge ($K_{{\rm c},i}$, pink squares), and their error bars obtained from the data in {\bf A} using $K_\nu=v_{\rm F}/v_\nu$. The blue dashed line is the non-interacting limit of these parameters $K_{\rm s,c}=1$. The index $i=1,2$ labels the first and the second subband, when the latter appear at $V_{\rm g}>-620$\,mV. The Fermi velocity of the second subband $v_{\rm F,2}$ is directly extracted from the data. The spin and charge velocities for the second subband, $v_{\rm c,2}$ and $v_{\rm s,2}$, are estimates only and their error bars are twice larger than for  $v_{\rm c,1}$ and $v_{\rm s,1}$.} 
    		\label{fig:velocity}
        \end{figure}
\begin{figure}
    		\centering
    		\includegraphics[width=\linewidth]{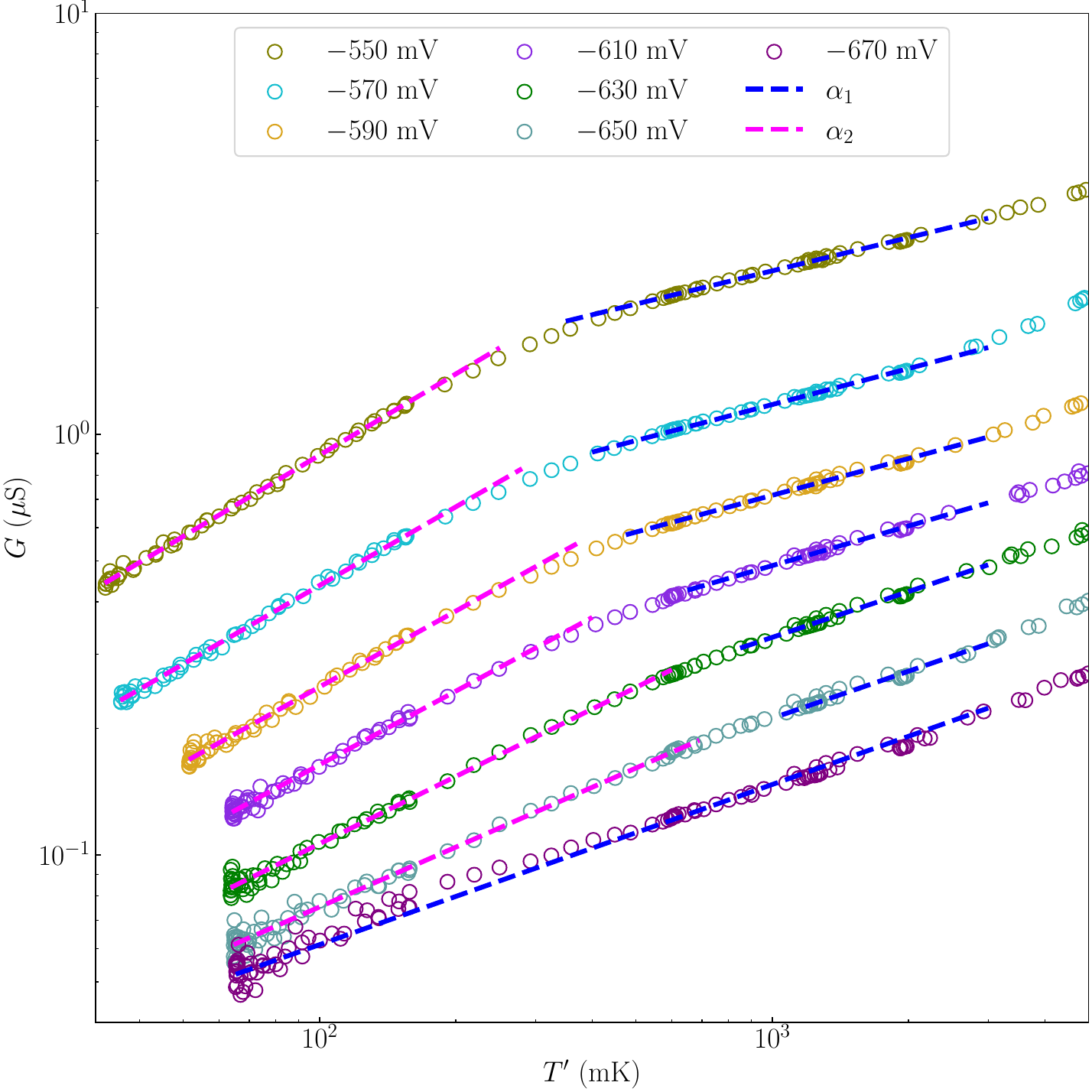}
    		\caption{The same data as in Fig.\ \ref{fig:conductance_exponents}A plotted as function of the effective temperature ${T^\prime=\sqrt[3]{T_0^3+T^3}}$. The saturation temperatures $T_0$ are different for different  gate voltages $V_{\rm g}$ given in the legend and the values of $T_0$ were taken from Fig.\ 5B. The blue and magenta dashed lines are the power-law fits giving the values of the exponents in Fig.\ 3B.
      %Extracted tunneling exponents $\alpha$ in the low (green) and high (black) - temperature range. 
      %The bulk ($\alpha_{bulk}$, blue) and the end ($\alpha_{end}$, red) tunneling exponent are calculated using Eq.\ (\ref{eq:alpha_bulk}) and Eq.\ (\ref{eq:alpha_end}) with the Luttinger parameters $K_{c,s}$ from Fig.\ \ref{fig:velocity}.
      }
    		\label{fig:fig_G0_Tprime}
        \end{figure}  

\newpage

\begin{figure}
    		\centering
    		\includegraphics[width=\linewidth]{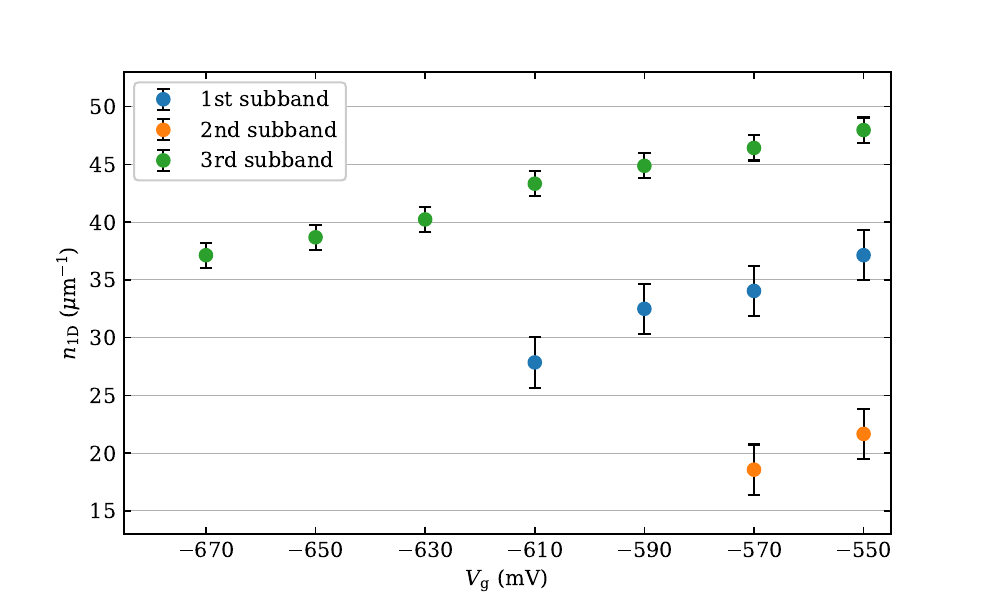}
    		\caption{The density of electrons in the wires $n_{\rm 1D}$ for the first three subbands as a function of wire-gate voltage $V_{\rm g}$. The data was extracted from the positions of the $B^\pm$ points in the spectroscopic maps with the error bars obtained as uncertainty in this extraction procedure, see details in the Spectroscopy subsection in the main text. 
            While we can track the density in the 1st subband across the whole range of gate-voltages, the visibility of the Fermi points of the 2nd and 3rd subband in the spectroscopy maps is limited at low densities, and can only be reliably determined starting from $V_{\rm g} = -610\,\rm mV$ (2nd subband) and $V_{\rm g} = -570\,\rm mV$ (3rd subband). However, it is very likely that they become occupied earlier than that, and their visibility is hindered by the ZBA, see the Discussion section in the main text.}
    		\label{fig:densities}
        \end{figure}  
\printbibliography[title={Supplementary References}]